\documentclass[
    11pt,
]{article}

\usepackage{authblk}

\usepackage[style=nature, natbib]{biblatex}

\PassOptionsToPackage{hyphens}{url}
\PassOptionsToPackage{dvipsnames,svgnames*,x11names*}{xcolor}

\usepackage{amsmath}                  

\usepackage{amssymb}                  

\usepackage{amsfonts}                 

\usepackage{bm}                       

\usepackage{calc}                     

\usepackage{cancel}                   

\usepackage{esint}                    

\usepackage{enumitem}                 

\usepackage{fancyhdr}                 

\usepackage[vario]{fancyref}          

\usepackage{graphicx}                 

\usepackage{parskip}                  

\usepackage{setspace}                 

\usepackage{xcolor}                   

\usepackage{hyperref}  

\usepackage{pgfplots}                 

\usepackage{geometry}                 

\usepackage{subfig}

\usepackage{caption}

\allowdisplaybreaks

\hypersetup{
    pdftitle={A minimal reaction-diffusion neural model generates C. elegans undulation},
    pdfauthor={Anshul Singhvi},
    colorlinks=true,
    linkcolor=Maroon,
    citecolor=Blue,
    urlcolor=Blue,
    breaklinks=true,
    pdfdisplaydoctitle = true
}

\usetikzlibrary{
    arrows.meta,
    calc,
    decorations,
    decorations.pathreplacing,
    decorations.footprints,
    math,
    patterns,
    shadows,
    external
}

\tikzset{>=stealth}

\pgfplotsset{compat=1.16}

\usepgfplotslibrary{
    polar,
    colormaps,
    colorbrewer,
    groupplots,
    statistics
}

\pgfplotscreateplotcyclelist{wong}{%
    color={rgb, 255 : red, 86  ; green, 180 ; blue, 233}, mark = *\\         
    color={rgb, 255 : red, 230 ; green, 159 ; blue, 0},   mark = square*\\   
    color={rgb, 255 : red, 0   ; green, 158 ; blue, 115}, mark = otimes*\\   
    color={rgb, 255 : red, 240 ; green, 228 ; blue, 66},  mark = star\\      
    color={rgb, 255 : red, 0   ; green, 114 ; blue, 178}, mark = diamond*\\  
    color={rgb, 255 : red, 213 ; green, 94  ; blue, 0},   mark = triangle*\\ 
    color={rgb, 255 : red, 204 ; green, 121 ; blue, 167}, mark = pentagon*\\ 
}

\pgfplotsset{every axis legend/.append style={%
        cells={anchor=west}
    },
    cycle list name = wong,
}

\tikzset{
  neuron/.style={
  circle,
  minimum size=6mm,
  very thick,
  draw=blue!50!black!50,
  top color=white,
  bottom color=blue!50!black!20, 
  font=\itshape,
  outer sep=2mm
  }
}

\usepackage[frozencache]{minted}

\begin{document}


\title{A minimal reaction-diffusion neural model generates {\emph{C. elegans}} undulation}

\author[1,2]{Anshul Singhvi}

\author[1]{Harold M. Hastings}

\author[3]{Jenny Magnes}

\author[3,4]{Susannah G. Zhang}

\affil[1]{Bard College at Simon's Rock}
\affil[2]{Columbia University}
\affil[3]{Vassar College}
\affil[4]{University of Georgia}

\date{\today}

\maketitle

\begin{abstract}

    The small (1 mm) nematode \emph{Caenorhabditis elegans} (\citet{corsi2015}, wormbook.org) has become widely used as a model organism; in particular the \emph{C. elegans} connectome has been completely mapped, and \emph{C. elegans} locomotion has been widely studied. We describe a minimal reaction-diffusion model for the locomotion of \emph{C. elegans}, using as a framework a simplified, stylized "descending pathway" of neurons as central pattern generator (CPG) (Xu \emph{et al.}, Proceedings of the National Academy of Sciences 115, 2018 \cite{xu2018}).  Finally, we realize a model of the required oscillations and coupling with a network of coupled Keener (IEEE Transactions on Systems, Man, and Cybernetics SMC-13, 1983 \cite{keener1983}) analog neurons.
    Note that Olivares \emph{et al.} (BioRxiv 710566, 2020 \cite{olivares2020}) present a likely more realistic model more distributed CPG. We use the simpler simulation to show that a small network of FitzHugh-Nagumo neurons (one of the simplest neuronal models) can generate key features of \emph{C. elegans} undulation, and thus locomotion, yielding a minimal, biomimetic model as a building block for further exploring \emph{C. elegans} locomotion.
\end{abstract}

\section{Introduction}\label{sec: intro}

The small (1 mm) nematode \emph{Caenorhabditis elegans} (\emph{C. elegans}) has become a widely used model organism (cf. http://www.wormbook.org \cite{corsi2015}), and has been among the most studied biological models of neuronal development and locomotion \citep{katz2016, corsi2015}.
The \emph{C. elegans} connectome has been completely mapped \citep{jabr} and, as described below, its locomotion has been widely studied (c.f. \citet{corsi2015, gjorgjieva2014, }).  There are a variety of neuronal models which can generate such undulation,
``When crawling on a solid surface, the nematode \emph{C. elegans} moves forward by propagating sinusoidal dorso-ventral retrograde contraction waves.  A uniform propagating wave leads to motion that undulates about a straight line.'' \citep{kim2011}.
A different type of locomotion, often called swimming, occurs when nematodes are submerged in a liquid medium. The nematodes “switch” between these two gaits, by changing the dynamics of the central pattern generator (CPG).

The purpose of this paper is to describe a minimal, biomimetic, reaction-diffusion model for the \emph{C. elegans} central pattern generator (CPG) \citep{xu2018, wen2012}.  We use simulation methods to show that a small network of \citet{fitzhugh1955}-\citet{nagumo1962} neurons, see also \cite{izhikevich2006fitzhugh} (one of the simplest neuronal models), based on a skeleton model of the \emph{C. elegans} CPG, can reproduce key features of \emph{C. elegans} undulation \citep{xu2018} \citep{magnes2012} \citep{magnes2020chaotic}, and thus locomotion.

Finally, we describe an analog electronic implementation of our model through solving a modified version of the FitzHugh-Nagumo neuron \cite{fitzhugh1955}, based upon an analog circuit originally proposed by \citet{keener1983}.  This circuit solves the Keener differential equations, and we adjusted it to allow diffusive coupling between neurons.  We constructed a small network with these ``neuro-mimetic'' circuits, and showed that their behaviour replicates FitzHugh-Nagumo simulated behaviour.

There are many other CPG models; for example,  \citet{olivares2020} proposes a distributed network of self-oscillating systems of neurons, instead of a structured chain like our proposed CPG.  Here we describe a minimal working model, rather than striving for fuller realism, in order to explore the fundamental components of a small CPG.  We aimed for a simple, "minimal" model to enable exploration of defects or other changes in the CPG in an efficient, reproducible and explainable way. Our model can thus be considered as intermediate between the early "neuro-mechanical" model of \citet{yuk2011shape} and more realistic models such as \cite{olivares2020}, a role similar to that of \citet{adamatzky2008} biochemical computation.

\section{The model central pattern generator}


Recall that central pattern generator is a small neural circuit which generates and regulates the movement of complex organisms.  This structure is present in different forms in many animals, and regulates many types of periodic motion. Changes in gait are driven by changes in the dynamics of the CPG, c.f. \citep{collins1994}. The CPG circuit topology is constant; the mode of locomotion depends on the sequence in which the neurons fire.

Our simple model \emph{C. elegans} central pattern generator has two principal components. The first is the \textbf{head oscillator}. As described by \citet{gjorgjieva2014}, the head oscillator consists of two “head neurons” with mutually inhibitory coupling. Oscillations are generated when this coupling destabilizes an excitable steady state. Here two Fitzhugh-Nagumo neurons, with oscillatory dynamics, stabilized $180^\circ$ out of phase by mutually inhibitory coupling. This provides a pair of out-of-phase stimuli which propagate through a \textbf{descending pathway} of pairs of coupled, excitable, dorsal and ventral neurons.  These follow the body of the worm, and are linked to motor neurons and muscles. The head oscillator drives the descending pathway, and the pathway is kept in sync by mutual inhibitory coupling between neurons.

We use 12 pairs of such neurons, as in \emph{C. elegans} coupled by model gap junctions. This coupling yields phase lags as we descend the pathway from head to tail, enabling the head oscillator to drive traveling waves of excitation along the body of the nematode. See  \Fref{fig: xu_cpg} for descending pathway of \citet{xu2018} and our simplified model; the latter depicted as a graph, wherein neurons are nodes, and the arrows between them represent connections.

\setcounter{figure}{0}
\begin{figure}[ht!]
    \centering
    \includegraphics[width=12.5cm]{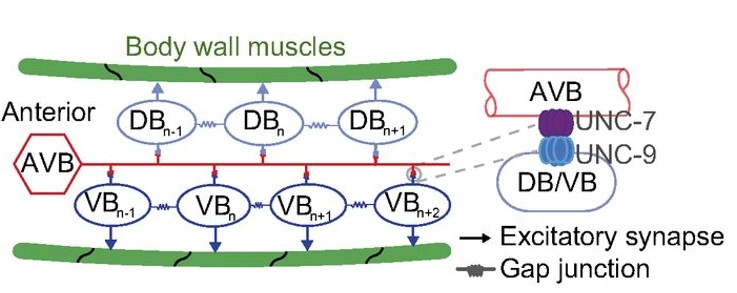}

\vspace{10mm}
    \centering
    \includegraphics [width=0.95\textwidth]{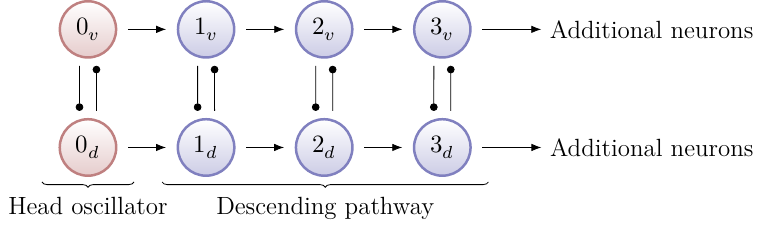}
    \caption{Top: The central pattern generator proposed by \citet{xu2018}. Reprinted from \cite{xu2018}, Figure 1A. Copyright 2018, National Academy of Sciences. Note the structure of the neurons, especially the head oscillator driving the descending pathway. “AVB-B gap junctions facilitate undulatory wave propagation during forward locomotion.” Bottom: Our simplified model, based upon the descending pathway of Xu. The model includes inhibitory connections between corresponding dorsal and ventral neurons (denoted by subscripts $d$ and $v$, respectively) and a descending pathway of gap junctions. We used chains of 6 to 12 such pairs of neurons have been  to model the descending pathways. Arrows indicate gap junctions; lines ending in solid circles indicate inhibitory synapses. Although not present in Xu’s model, inhibitory connections below the head oscillator appear in \citet{chen2006wiring}, \citet{boyle2012gait} and \citet{zhen2015c}.}
    \label{fig: xu_cpg}
\end{figure}

\section{The FitzHugh-Nagumo Neuronal Model}\label{sec: fhn}

As described above, we sought to use the simplest relevant neuronal model.  The classical Hodgkin-Huxley\cite{hodgkin1952} model of squid neurons has led to a variety of simpler conduction models, including the Morris-Lecar\cite{morris1981} and \citet{fitzhugh1955}-\citet{nagumo1962} (FHN) models. The FHN model consists of two dynamical variables; a fast activator variable $v$ corresponding to the (rescaled) membrane potential, and a slow inhibitor variable $w$ corresponding to a generalized gating variable.

\begin{equation}
    \label{eq: fhn}
    \begin{aligned}
        \frac{dv}{dt} &= f(v) - w  + I_\mathrm{ext}\\
        \frac{dw}{dt} &= \varepsilon(v - \gamma w + \alpha)\\
        f(v) &=  v - \frac{v^3}{3}
    \end{aligned}
\end{equation}

The parameter $I_\mathrm{ext}$ is an external driving current, and is used here to model the effect of gap junctions and synapses upon membrane potential.  The parameter $\alpha$ determines the vertical position of the $w$-nullcline and thus controls excitability. Action potentials can also be generated by a current injection corresponding to $I_\mathrm{ext}$. Finally, the parameter $\varepsilon$ determines ratio between the time scales of the fast activator variable $v$ and the slow inhibitor variable $w$; $\gamma$ determines the effect of the growth of the slow gate variable $w$. We used the following standard parameter values: $\varepsilon = 0.08, \gamma = 0.8$ \cite{izhikevich2006fitzhugh}. FitzHUgh-Nagumo neurons can display either (self-)oscillatory or excitable dynamics; given a sufficiently large stimulus (here a pulse of injected current $I$), an excitable neuron generates an action potential. An oscillatory neuron generates a regular sequence of action potentials.  In comparison to the standard value $\alpha = 0.7$ \cite{izhikevich2006fitzhugh}, we adjusted the parameter $\alpha$ to control neuronal dynamics, with values larger than the oscillatory-excitable boundary $\alpha_0 \approx 0.467$ generating excitable dynamics from a stable steady state, and smaller values generating oscillatory dynamics as the steady state is destabilized.

Moreover, $f(v)$ can be any function which retains the appropriate dynamics, in that it has the same general shape as the cubic $f(v) = \frac{v^3}{3} - v$.  For example, $f(v)$ could be replaced by the cubic-like I-V curve of the tunnel diode in the \citet{nagumo1962} circuit, or even the piecewise linear approximation generated in the \citet{keener1983} circuit used in our implementation.

\begin{figure}[ht!]
    \centering
    \includegraphics[width=12.5cm]{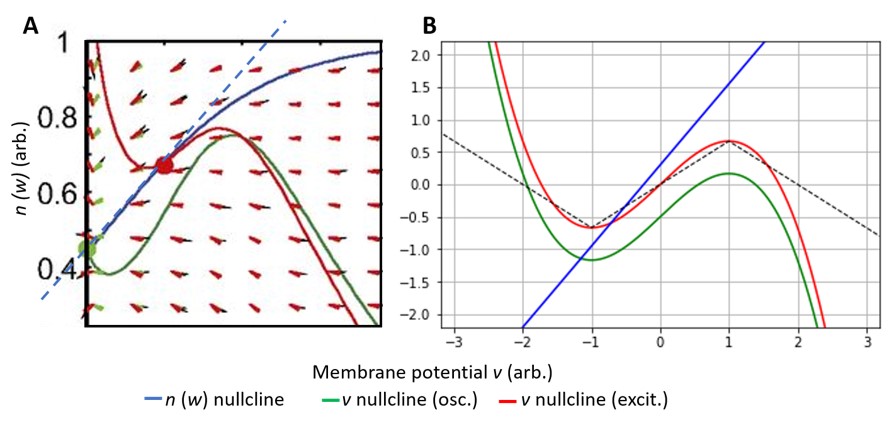}
    \caption{Dynamics of (A) \textit{C. elegans} neurons adapted from Xu \textit{et al.}, Fig. 5B, Copyright National Academy of Sciences \cite{xu2018} (left) and (B) Fitzhugh-Nagumo neurons \cite{fitzhugh1955, nagumo1962} (right). We have identified Xu's $n$-variable (the potassium current in the Hodgkin-Huxley model \cite{hodgkin1952} ( with the $w$-variable (slow inhibitor) in the Fitzhugh-Nagumo model. In both models, the nullclines for sodium dynamics ($v$) has a cubic-like shape. We show the $v$-nullcline in two positions: one generating oscillatory dynamics (green); the other excitable (non-oscillatory) dynamics (green). The type of dynamics, oscillatory or excitable, is controlled by coupling of AVB-B gap junctions in Xu's model. The type of dynamics can be controlled by the injected current $I$, an analog of the above AVB-B coupling or the parameter $\alpha$ in the Fitzhugh-Nagumo model. We added a dashed blue line to (A) Xu's experimentally derived simulation (A) to indicate that the $n$-nullcline is essentially linear in the relevant range of dynamics. It therefore seems reasonable to use the linear $w$-nullcline in the simpler Fitzhugh-Nagumo model. Finally, the thin dashed black line in the graph of Fitzhugh-Nagumo nullclines (A) represents the dynamics of Keener's \cite{keener1983} implementation of the Nagumo circuit (below). The Keener circuit yields a piecewise linear approximation to  Fitzhugh-Nagumo $v$ dynamics.}
    \label{fig: nullclines}
\end{figure}

\citet{xu2018} described a simplified two-variable model for \emph{C. elegans} neurons, consisting of a fast, cubic-like activator variable (see the $v$-nullcline in \Fref{fig: nullclines}) and a slow, non-linear inhibitor variable (see the $n$-nullcline). The \citet{xu2018} neuronal model can thus be interpreted as a Fitzhugh-Nagumo type model neuron.


Following \citet{collins1994,yanagita2005}, we now use
generalized diffusion coupling between FHN neurons to model gap junctions and synapses. The effect of gap junctions and synapses upon the membrane potential is modeled by replacing the external current $I_\mathrm{ext}$ of the system, by a generalized diffusion term, $D \max(\Delta v,0)$. A positive diffusion coefficient $D$ is used to simulate a gap junction, or an excitatory synapse and a negative coefficient to simulate an inhibitory synapse (\cite{collins1994}).

This yields  a system of diffusion-coupled FitzHugh-Nagumo equations, c.f. \citet{essaki2015chimeric}:

\begin{equation}
    \label{eq: fhnd}
    \begin{aligned}
        \frac{dv}{dt}   &= f(v) - w  + \color{black}{D(\Delta v)}\\
        \frac{dw}{dt}   &= \varepsilon (v - \gamma w + \alpha)\\
        f(v) &= v - \frac{v^3}{3}
    \end{aligned}
\end{equation}

where $\Delta v$ is the rectified difference in voltage between the driving and driven neurons, essentially $\Delta v = max(v_{driven} - v_{driving},0)$. For example, consider the potential of an oscillatory FitzHugh-Nagumo neuron to drive an excitable FitzHugh-Nagumo neuron. The consequent dynamics depends upon the magnitude of positive diffusion coupling $D > 0$: $D$ must be sufficiently large to generate an action potential in the driven excitable neuron, which then responds with a time delay correspodning inversely to the magnitude of $D$.

\section{Simulation}

%

We start with the network shown in \Fref{fig: xu_cpg}, bottom. We perform  simulations in Python, using the standard SciPy ODE solvers, which wrap LSODA, the Livermore Solver for Ordinary Differential Equations. The following block of code illustrates these calculations. Much like \citet{izquierdo2018} did, we take a segment-based approach to the worm model.  Pairs of muscles on either side determine the overall angular displacement per segment.  The magnitude of the “contraction” resulting from the smoothing is interpreted as an angular displacement for that segment.  We use cubic spline interpolation to smooth the worm body.

\small
\begin{minted}{python}

#Imports
import numpy as np
from scipy.integrate import odeint
import matplotlib.pyplot as plt
import math

#constants
# d for dorsal
# v for ventral
# x for FHN membrane potential, namely v in equations (1)-(2)
# x for FHN gate variable, namely w in equations (1)-(2)

# initial data, some symmetry breaking
# dynamics largely independent of intial data
dx0= 1.0
dy0=-0.51
vx0= 1
vy0= -0.49
# more, omitted here


#vector of x and y values
v=[vx0,vy0,dx0,dy0,vx1,vy1,dx1,dy1,vx2, ...]

# time range for simulations (arbitrary units)
t = np.linspace(0, 2000, 20001)

def FHND(v, t):
    # calculates the derivatives in FHN with diffusion
    vx0, vy0, dx0, dy0, vx1, vy1, dx1, dy1, vx2, vy2, ... = v
    # v denotes vector passed to the function FHND
    # FitzHugh-Nagumo parameters, see eqs. (2) and (3).
    # allow for different epsilons
    e0 = 0.08          # epsilon for head neurons
    e1 = 0.08          # epsilon for body neurons
    g = 0.8
    b0 = 0.46
    b1 = 0.47
    # diffusion constants
    Dhead = -0.2
    # Drest= -0.02
    Drest= 0
    Dgap=0.05
    J = 0

    dvdt=[
         # neurons 0 (ventral and dorsal)
        vx0-(vx0**3/3)- vy0+ Dhead*max(dx0-vx0,0)                  + J,
        e0*(vx0-g*vy0+b0),
        dx0-(dx0**3/3)- dy0+ Dhead*max(vx0-dx0,0)                  + J,
        e0*(dx0-g*dy0+b0),
        # coupled by 1-way diffusion
        # negative diffusion constant Dhead
        # simulated inhibitory synapse

        # neurons 1 (ventral and dorsal)
        vx1-(vx1**3/3)- vy1+ Drest*max(dx1-vx1,0) + Dgap*max(vx0-vx1,0) + J,
        e1*(vx1-g*vy1+b1),
        # driven by neurons 0 through one-way
        dx1-(dx1**3/3)- dy1+ Drest*max(vx1-dx1,0) + Dgap*max(dx0-dx1,0) + J,
        e1*(dx1-g*dy1+b1),
        # Dgap simulates gap junction in descending chain
        # Drest simulates inhibitary synapse
        # coupling ventral and dorsal neurons 1, ...
          # more, omitted
    return dvdt
\end{minted}

\normalsize

Muscle response is simulated by applying Gaussian filters to the positive component of membrane potentials:

\small
\begin{minted}{python}
import scipy.ndimage.filters as filt
effective_signal=np.zeros((20001,24))
for neuron_number in range(0,number_of_neurons):
    effective_signal[:,neuron_number]=
    (sol[:,4*neuron_number]+abs(sol[:,4*neuron_number]))/2
    effective_signal[:,neuron_number]+=
         (sol[:,4*neuron_number+2]+abs(sol[:,4*neuron_number+2]))/2
    effective_signal[:,neuron_number]+=
         filt.gaussian_filter1d(effective_signal[:,neuron_number],40)
    # Gaussian filter simulates a combination of
    # muscle response to neural stimulus,
    # effects of elastic properties of worm and
    # also interaction with fluid
\end{minted}
\normalsize

\begin{figure}[ht!]
    \centering
    \includegraphics[width=0.95\textwidth]{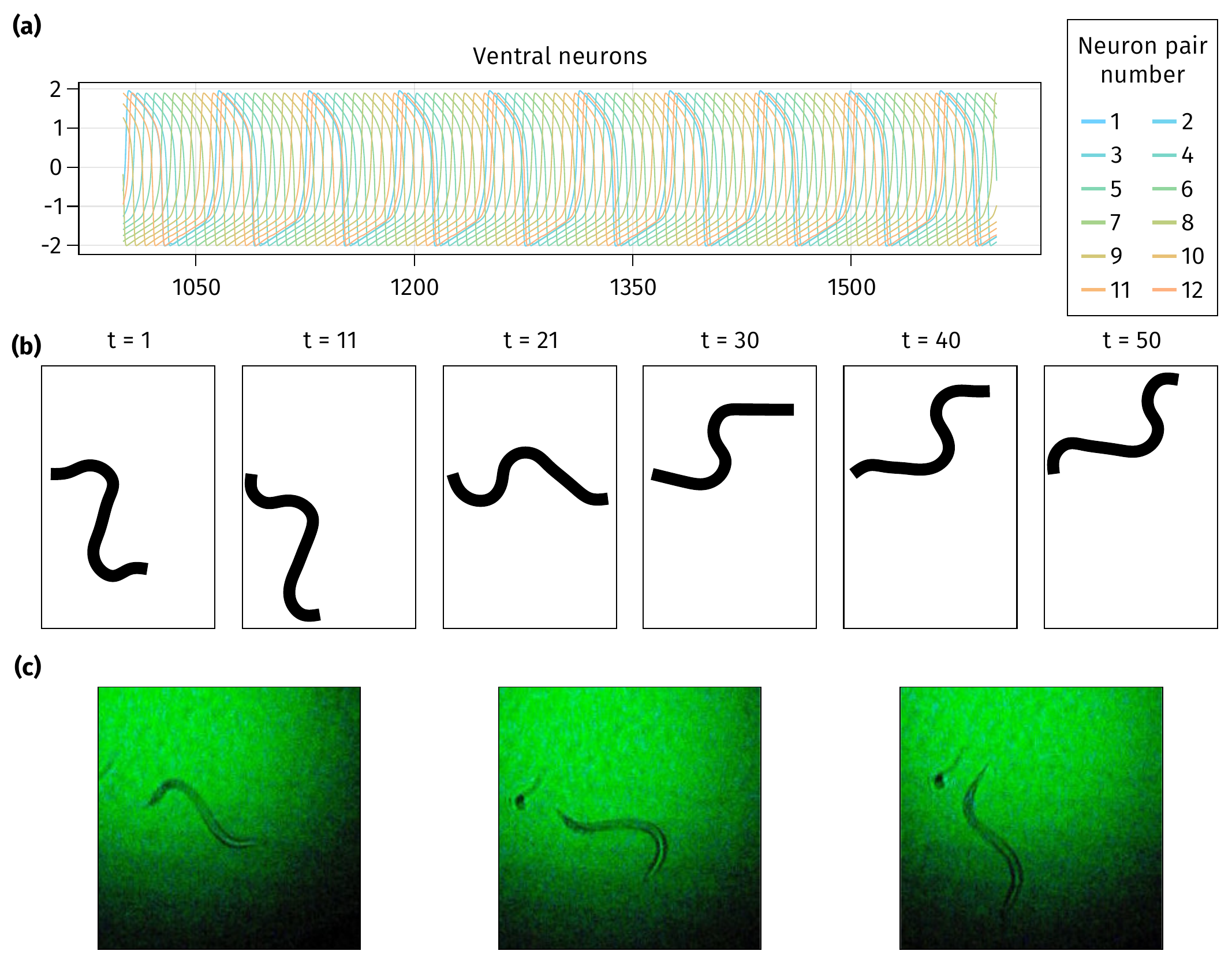}
    \caption{A "dashboard view" of our simulated \emph{C. elegans}.  (a) Graphical representation of a traveling wave of simulated membrane potentials of ventral neurons, numbered from head to tail. Time and membrane potential units are arbitrary.  (b) Successive "images" of the simulated \emph{C. elegans}. Muscle response to neural potentials was simulated using a Gaussian filter to simulate muscle dampening, one dorsal (resp., ventral) muscle per dorsal (resp., ventral) neuron. A cubic spline is used to simulate a "smoothed" worm body between joints as nodes. (c) For comparison, here are shadow images of a real \emph{C. elegans} in a cuvette.}
    \label{fig: worm_neuron_dash}
\end{figure}

The worm is then described initially as a sequence of nodes, joined by line segments, with the angle at each node determined by muscular tension. This is followed by a cubic spline interpolation. Finally, we generated a simple video  by fixing the head of a worm to the origin; see \Fref{fig: worm_neuron_dash}.

\emph{Simulation versus experimental results.} Our model generates traveling waves of approximately sinusoidal form, with approximately constant amplitude from head to tail. In comparison, real \emph{C. elegans} undulations are characterized by approximately sinusoidal oscillations, but with amplitude decreasing slightly from head to tail, c.f. \cite{xu2018}. In addition, a search for "markers of chaos" as in \citet{magnes2020chaotic} finds that undulations of the simulated worm are too regular (neutrally stable), in contrast to unduations of real nematodes (with $\mathrm{Max}\left\{Re(\lambda)\right\} \approx 1 \mathrm{s}^{-1}$), likely because we have yet to implement proprioceptive neuronal inputs, c.f. \citet{calhoun2014maximally} and references therein. The neutral stability of our model may allow it to respond readily but relatively consistently to inputs. Moreover, a small, negative $Max \{Re(\lambda)\}$ would typically maintain relatively consistent undulation, but be readily overcome by somewhat larger inputs or noise, the latter generating random changes in undulation pattern.

Further comparisons and a more detailed non-linear analysis of real \emph{C. elegans} undulation will appear in a forthcoming paper [Zhang \emph{et al.}, in preparation].

\section{Analog implementation}

\begin{figure}[ht!]
    \centering
    \includegraphics[width=0.95\textwidth]{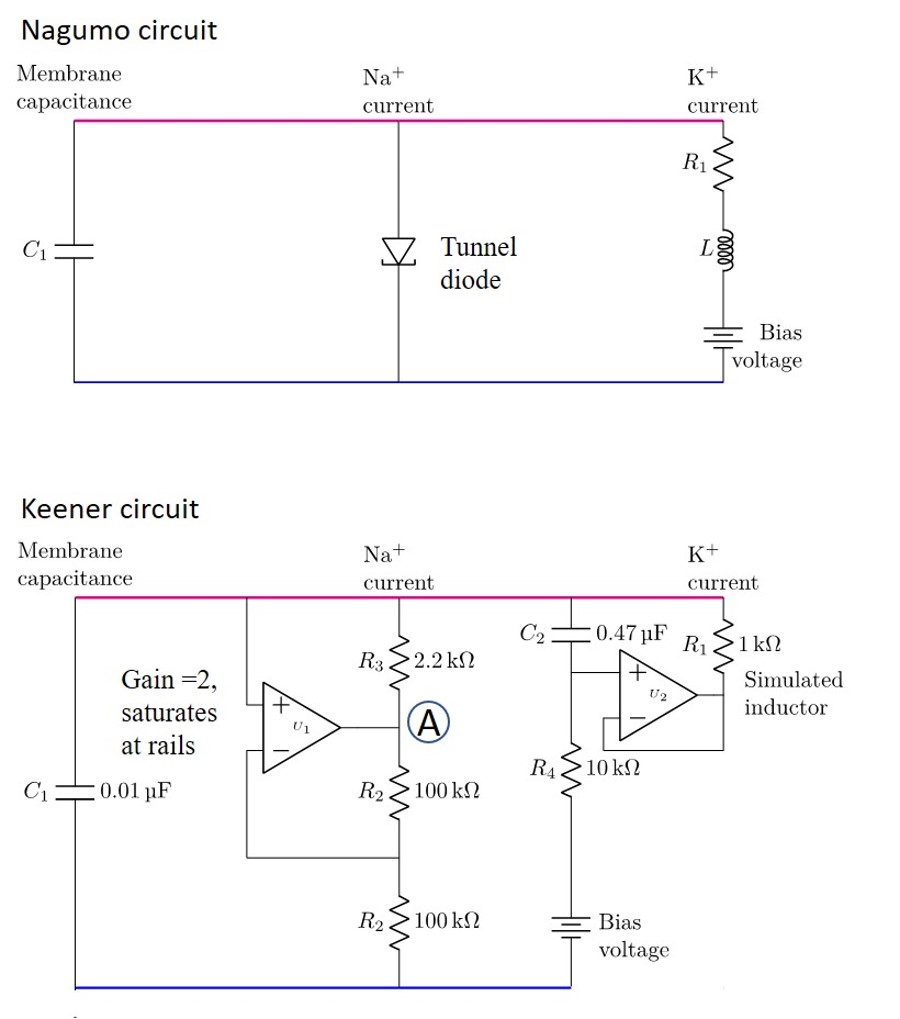}
    \caption{Top: The original circuit proposed by \citet{nagumo1962}.  Note the use of inductors for differentiation, and a tunnel diode to supply a cubic-like V-I curve for sodium dynamics in the Fitzhugh-Nagumo model. Bottom: Our circuit, a minor modification of \citet{keener1983}, using a single pair of power supplies and fine bias control, is shown in the same layout, to make the similarities and differences more explicit. The cubic-like V-I curve for sodium dynamics is generated by op-amp $U_1$ and associated resistors. In fact this circuit provides a polygonal approximation to  Fitzhugh-Nagumo sodium dynamics. The inductor in the Nagumo circuit is simuated by op-amp $U_2$ and associated passive components. Roughly, differentiation is simulated by "anti-integration."}
    \label{fig: analog_neurons}
\end{figure}

We show that a diffusion-coupled \citet{keener1983} electronic analog neurons can imitate key features of the dynamics of our network of Fitzhugh-Nagumo neurons. Recall that \citet{nagumo1962} proposed a circuit to simulate a FitzHugh-Nagumo neuron, shown in \Fref{fig: analog_neurons}.  It used a tunnel diode to achieve a cubic-like activation function, and an inductor to differentiate the potassium current (slow gate variable)  \Fref{fig: analog_neurons}

However, given that tunnel diodes are expensive and rarely available, \citet{keener1983} proposed a modified Nagumo circuit which used the saturation properties of operational amplifiers ("op-amps") to achieve cubic-like non-linearity in the FHN model. This yields a piecewise linear approximation to the Fitzhugh-Nagumo $v$ (sodium, fast activator) nullclines, as shown in \Fref{fig: nullclines}. The nullclines are sufficiently similar that the dynamics are effectively the same \cite{keener1983}. Keener also used an op-amp and a capacitor to simulate the inductor in the original Nagumo circuit. Roughly, differentiation is simulated by "anti-integration.Later \citet{sanchez1989circuit} proposed a CMOS implementation of the Nagumo circuit.

\begin{figure}[ht!]
    \centering
    \pgfplotsset{}
    \includegraphics[width=0.95\textwidth]{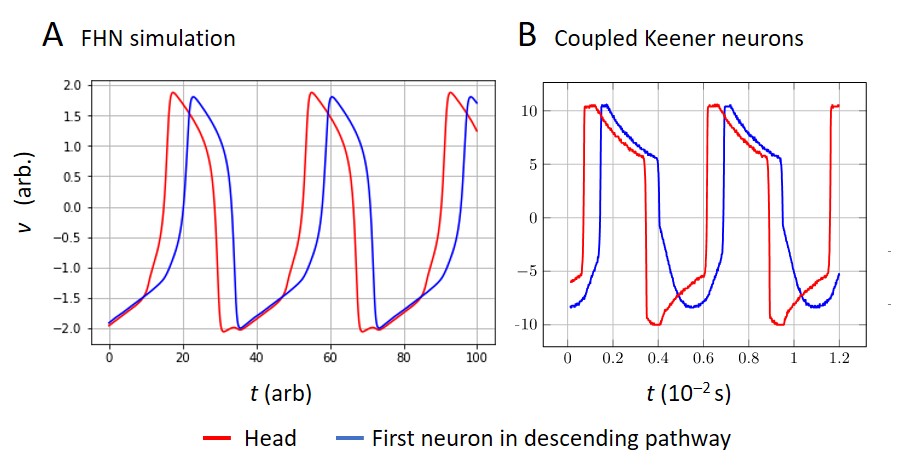}
    \caption{Dynamics of analog implementation of coupling from one neuron in our head oscillator to the next corresponding neuron in our descending chain of neurons; see \Fref{fig: xu_cpg} . (A) Theoretical results, as in \Fref{fig: worm_neuron_dash}. (B) Experimental results from diffusion coupled Keener neurons. Note similarity with (A). Our experimental period, $\sim 5 ms$ agrees with that of \citet{keener1983}. The head oscillator and additional downstream couplings are also similarly reproduced by diffusion-coupled Keener neurons.  }
    \label{fig: Keener_simulation}
\end{figure}

Finally, diffusion coupling is simulated by a follower, and then a coupling resistor followed by a diode (positive diffusion, excitatory synapse or gap junction) or a follower, inverting amplifier of unit gain, coupling resistor and a diode in series (negative diffusion, inhibitory synapse), c.f. \cite{collins1994,yanagita2005} for the role of diffusion, c.f. \citet{essaki2015chimeric} for the use of resistors to emulate diffusion. The diffusion constant is inversely proportional to the coupling resistance. Coupling resistors are tuned to obtain desired synchronization or time delays; for example, coupling resistors in the descending chain of neurons are $\sim$ 1-3 M$\Omega$, with larger resistors yielding longer delays until eventually the injected current is too small to generate an action potential. Typical experimental results for a pair of neurons in the descending chain coupled by a gap junction are shown in \Fref{fig: Keener_simulation}.

\newpage
\section{Conclusion}

We have shown that the undulatory motion of \emph{C. elegans} can be generated using a simple network of Fitzhugh-Nagumo neurons, the network capturing  using a structured central pattern generator, and simple, biomimetic neurons. In particular, our simple, "minimal" model generates traveling waves of sinusoidal undulations, as in \cite{xu2018}. An initial exploration shows that small changes in parameters do not alter qualitative dynamics; further study is needed to explore the range of possible dynamics.

Our model thus lies in between the "Shape memory alloy-based small crawling robot" mechanical model of \citet{yuk2011shape} and more realistic (and complex) descriptions of \citet{izquierdo2018} and \citet{olivares2020}, in the spirit of \citet{adamatzky2008} description of the neural system as a computational system. Because of the simplicity and flexibility our our model, we expect that it may prove useful is studying the effects of stimuli, aging and mutations upon the dynamics of \emph{C. elegans} undulation and locomotion.

\emph{Acknowledgement.} We acknowledge assistance from our students Cheris Congo, Miranda Hulsey-Vincent, Rifah Tasnim and Naol Negassa.

\newpage
\section{References}

\printbibliography[heading=none]

@article{adamatzky2008,
  title = {Universal Computation with Limited Resources: {{Belousov}}-{{Zhabotinsky}} and {{Physarum}} Computers},
  shorttitle = {{{UNIVERSAL COMPUTATION WITH LIMITED RESOURCES}}},
  author = {Adamatzky, Andrew and De Lacy Costello, Ben and Shirakawa, Tomohiro},
  date = {2008-08},
  journaltitle = {International Journal of Bifurcation and Chaos},
  shortjournal = {Int. J. Bifurcation Chaos},
  volume = {18},
  pages = {2373--2389},
  issn = {0218-1274, 1793-6551},
  doi = {10.1142/S0218127408021750},
  url = {https://www.worldscientific.com/doi/abs/10.1142/S0218127408021750},
  urldate = {2020-03-26},
  langid = {english},
  number = {08}
}

@article{boyle2012gait,
  title={Gait modulation in C. elegans: an integrated neuromechanical model},
  author={Boyle, Jordan Hylke and Berri, Stefano and Cohen, Netta},
  journal={Frontiers in computational neuroscience},
  volume={6},
  pages={10},
  year={2012},
  publisher={Frontiers}
}

@article{calhoun2014maximally,
  title={Maximally informative foraging by Caenorhabditis elegans},
  author={Calhoun, Adam J and Chalasani, Sreekanth H and Sharpee, Tatyana O},
  journal={Elife},
  volume={3},
  pages={e04220},
  year={2014},
  publisher={eLife Sciences Publications Limited}
}

@article{chen2006wiring,
  title={Wiring optimization can relate neuronal structure and function},
  author={Chen, Beth L and Hall, David H and Chklovskii, Dmitri B},
  journal={Proceedings of the National Academy of Sciences},
  volume={103},
  number={12},
  pages={4723--4728},
  year={2006},
  publisher={National Acad Sciences}
}

@article{collins1994,
  title = {Hard-Wired Central Pattern Generators for Quadrupedal Locomotion},
  author = {Collins, J. J. and Richmond, S. A.},
  date = {1994-09},
  journaltitle = {Biological Cybernetics},
  shortjournal = {Biol. Cybern.},
  volume = {71},
  pages = {375--385},
  issn = {0340-1200, 1432-0770},
  doi = {10.1007/BF00198915},
  url = {http://link.springer.com/10.1007/BF00198915},
  urldate = {2020-03-11},
  langid = {english},
  number = {5}
}

@article{corsi2015,
  title = {A {{Transparent}} Window into Biology: {{A}} Primer on {{Caenorhabditis}} Elegans},
  shorttitle = {A {{Transparent}} Window into Biology},
  author = {Corsi, Ann K.},
  date = {2015-06-18},
  journaltitle = {WormBook},
  shortjournal = {WormBook},
  pages = {1--31},
  issn = {15518507},
  doi = {10.1895/wormbook.1.177.1},
  url = {http://www.wormbook.org/chapters/www_celegansintro/celegansintro.html},
  urldate = {2020-03-21}
}

@article{essaki2015chimeric,
  title={A chimeric path to neuronal synchronization},
  author={Essaki Arumugam, Easwara Moorthy and Spano, Mark L},
  journal={Chaos: An Interdisciplinary Journal of Nonlinear Science},
  volume={25},
  number={1},
  pages={013107},
  year={2015},
  publisher={AIP Publishing LLC}
}

@article{fitzhugh1955,
  title = {Mathematical Models of Threshold Phenomena in the Nerve Membrane},
  author = {FitzHugh, Richard},
  date = {1955-12},
  journaltitle = {The Bulletin of Mathematical Biophysics},
  shortjournal = {Bulletin of Mathematical Biophysics},
  volume = {17},
  pages = {257--278},
  issn = {0007-4985, 1522-9602},
  doi = {10.1007/BF02477753},
  url = {http://link.springer.com/10.1007/BF02477753},
  urldate = {2020-03-11},
  langid = {english},
  number = {4}
}

@article{gjorgjieva2014,
  title = {Neurobiology of {{Caenorhabditis}} Elegans {{Locomotion}}: {{Where Do We Stand}}?},
  shorttitle = {Neurobiology of {{Caenorhabditis}} Elegans {{Locomotion}}},
  author = {Gjorgjieva, Julijana and Biron, David and Haspel, Gal},
  date = {2014-06-01},
  journaltitle = {BioScience},
  volume = {64},
  pages = {476--486},
  issn = {1525-3244, 0006-3568},
  doi = {10.1093/biosci/biu058},
  url = {http://academic.oup.com/bioscience/article/64/6/476/289633/Neurobiology-of-Caenorhabditis-elegans-Locomotion},
  urldate = {2020-03-22},
  langid = {english},
  number = {6}
}

@article{hodgkin1952,
  title = {A Quantitative Description of Membrane Current and Its Application to Conduction and Excitation in Nerve},
  author = {Hodgkin, A. L. and Huxley, A. F.},
  date = {1952-08-28},
  journaltitle = {The Journal of Physiology},
  shortjournal = {The Journal of Physiology},
  volume = {117},
  pages = {500--544},
  issn = {0022-3751, 1469-7793},
  doi = {10.1113/jphysiol.1952.sp004764},
  url = {https://onlinelibrary.wiley.com/doi/abs/10.1113/jphysiol.1952.sp004764},
  urldate = {2020-03-23},
  langid = {english},
  number = {4}
}

@article{izhikevich2006fitzhugh,
  title={Fitzhugh-nagumo model},
  author={Izhikevich, Eugene M and FitzHugh, Richard},
  journal={Scholarpedia},
  volume={1},
  number={9},
  pages={1349},
  year={2006}
}

@article{izquierdo2018,
  title = {From Head to Tail: A Neuromechanical Model of Forward Locomotion in {{{\emph{Caenorhabditis}}}} {\emph{Elegans}}},
  shorttitle = {From Head to Tail},
  author = {Izquierdo, Eduardo J. and Beer, Randall D.},
  date = {2018-10-19},
  journaltitle = {Philosophical Transactions of the Royal Society B: Biological Sciences},
  shortjournal = {Phil. Trans. R. Soc. B},
  volume = {373},
  pages = {20170374},
  issn = {0962-8436, 1471-2970},
  doi = {10.1098/rstb.2017.0374},
  url = {https://royalsocietypublishing.org/doi/10.1098/rstb.2017.0374},
  urldate = {2020-03-11},
  langid = {english},
  number = {1758}
}

@online{jabr,
  title = {The {{Connectome Debate}}: {{Is Mapping}} the {{Mind}} of a {{Worm Worth It}}?},
  shorttitle = {The {{Connectome Debate}}},
  author = {Jabr, Ferris},
  url = {https://www.scientificamerican.com/article/c-elegans-connectome/},
  urldate = {2020-03-21},
  langid = {english},
  organization = {{Scientific American}}
}

@article{katz2016,
  title = {Evolution of Central Pattern Generators and Rhythmic Behaviours},
  author = {Katz, Paul S.},
  date = {2016-01-05},
  journaltitle = {Philosophical Transactions of the Royal Society B: Biological Sciences},
  shortjournal = {Philosophical Transactions of the Royal Society B: Biological Sciences},
  volume = {371},
  pages = {20150057},
  publisher = {{Royal Society}},
  doi = {10.1098/rstb.2015.0057},
  url = {https://royalsocietypublishing.org/doi/10.1098/rstb.2015.0057},
  urldate = {2020-03-21},
  number = {1685}
}

@article{keener1983,
  title = {Analog Circuitry for the van Der {{Pol}} and {{FitzHugh}}-{{Nagumo}} Equations},
  author = {Keener, James P.},
  date = {1983-09},
  journaltitle = {IEEE Transactions on Systems, Man, and Cybernetics},
  shortjournal = {IEEE Trans. Syst., Man, Cybern.},
  volume = {SMC-13},
  pages = {1010--1014},
  issn = {0018-9472, 2168-2909},
  doi = {10.1109/TSMC.1983.6313098},
  url = {http://ieeexplore.ieee.org/document/6313098/},
  urldate = {2020-03-11},
  number = {5}
}

@article{kim2011,
  title = {The Shallow Turn of a Worm},
  author = {Kim, D. and Park, S. and Mahadevan, L. and Shin, J. H.},
  date = {2011-05-01},
  journaltitle = {Journal of Experimental Biology},
  shortjournal = {Journal of Experimental Biology},
  volume = {214},
  pages = {1554--1559},
  issn = {0022-0949, 1477-9145},
  doi = {10.1242/jeb.052092},
  url = {http://jeb.biologists.org/cgi/doi/10.1242/jeb.052092},
  urldate = {2020-03-21},
  langid = {english},
  number = {9}
}

@article{magnes2012,
  title = {Quantitative {{Locomotion Study}} of {{Freely Swimming Micro}}-Organisms {{Using Laser Diffraction}}},
  author = {Magnes, Jenny and Susman, Kathleen and Eells, Rebecca},
  date = {2012-10-25},
  journaltitle = {Journal of Visualized Experiments},
  shortjournal = {JoVE},
  pages = {4412},
  issn = {1940-087X},
  doi = {10.3791/4412},
  url = {http://www.jove.com/video/4412/quantitative-locomotion-study-freely-swimming-micro-organisms-using},
  urldate = {2020-03-21},
  langid = {english},
  number = {68}
}

@article{magnes2020chaotic,
  title={Chaotic markers in dynamic diffraction},
  author={Magnes, Jenny and Hastings, Harold and Hulsey-Vincent, Miranda and Congo, Cheris and Raley-Susman, Kathleen and Singhvi, Anshul and Hatch, Tyler and Szwed, Erik},
  journal={Applied Optics},
  volume={59},
  number={22},
  pages={6642--6647},
  year={2020},
  publisher={Optical Society of America}
}

@article{morris1981,
  title = {Voltage Oscillations in the Barnacle Giant Muscle Fiber},
  author = {Morris, C. and Lecar, H.},
  date = {1981-07},
  journaltitle = {Biophysical Journal},
  shortjournal = {Biophysical Journal},
  volume = {35},
  pages = {193--213},
  issn = {00063495},
  doi = {10.1016/S0006-3495(81)84782-0},
  url = {https://linkinghub.elsevier.com/retrieve/pii/S0006349581847820},
  urldate = {2020-03-23},
  langid = {english},
  number = {1}
}

@article{nagumo1962,
  title = {An {{Active Pulse Transmission Line Simulating Nerve Axon}}},
  author = {Nagumo, J. and Arimoto, S. and Yoshizawa, S.},
  date = {1962-10},
  journaltitle = {Proceedings of the IRE},
  shortjournal = {Proc. IRE},
  volume = {50},
  pages = {2061--2070},
  issn = {0096-8390},
  doi = {10.1109/JRPROC.1962.288235},
  url = {http://ieeexplore.ieee.org/document/4066548/},
  urldate = {2020-03-11},
  number = {10}
}

@article{olivares2020,
  title={A neuromechanical model of multiple network rhythmic pattern generators for forward locomotion in C. elegans},
  author={Olivares, Erick and Izquierdo, Eduardo and Beer, Randall},
  journal={BioRxiv},
  pages={710566},
  year={2020},
  publisher={Cold Spring Harbor Laboratory}
}

@inproceedings{sanchez1989circuit,
  title={Circuit implementation of neural FitzHugh-Nagumo equations},
  author={Sanchez-Sinencio, E and Linares-Barranco, Bernab},
  booktitle={Proceedings of the 32nd Midwest Symposium on Circuits and Systems,},
  pages={244--247},
  year={1989},
  organization={IEEE}
}

@article{wen2012,
  title = {Proprioceptive {{Coupling}} within {{Motor Neurons Drives C}}. Elegans {{Forward Locomotion}}},
  author = {Wen, Quan and Po, Michelle D. and Hulme, Elizabeth and Chen, Sway and Liu, Xinyu and Kwok, Sen Wai and Gershow, Marc and Leifer, Andrew M. and Butler, Victoria and Fang-Yen, Christopher and Kawano, Taizo and Schafer, William R. and Whitesides, George and Wyart, Matthieu and Chklovskii, Dmitri B. and Zhen, Mei and Samuel, Aravinthan D.T.},
  date = {2012-11},
  journaltitle = {Neuron},
  shortjournal = {Neuron},
  volume = {76},
  pages = {750--761},
  issn = {08966273},
  doi = {10.1016/j.neuron.2012.08.039},
  url = {https://linkinghub.elsevier.com/retrieve/pii/S0896627312008057},
  urldate = {2020-03-21},
  langid = {english},
  number = {4}
}

@article{xu2018,
  title = {Descending Pathway Facilitates Undulatory Wave Propagation in {{{\emph{Caenorhabditis}}}}{\emph{ Elegans}} through Gap Junctions},
  author = {Xu, Tianqi and Huo, Jing and Shao, Shuai and Po, Michelle and Kawano, Taizo and Lu, Yangning and Wu, Min and Zhen, Mei and Wen, Quan},
  date = {2018-05-08},
  journaltitle = {Proceedings of the National Academy of Sciences},
  shortjournal = {Proc Natl Acad Sci USA},
  volume = {115},
  pages = {E4493-E4502},
  issn = {0027-8424, 1091-6490},
  doi = {10.1073/pnas.1717022115},
  url = {http://www.pnas.org/lookup/doi/10.1073/pnas.1717022115},
  urldate = {2020-03-11},
  langid = {english},
  number = {19}
}

@article{yanagita2005,
  title = {Pair of Excitable {{FitzHugh}}-{{Nagumo}} Elements: {{Synchronization}}, Multistability, and Chaos},
  shorttitle = {Pair of Excitable {{FitzHugh}}-{{Nagumo}} Elements},
  author = {Yanagita, T. and Ichinomiya, T. and Oyama, Y.},
  date = {2005-11-28},
  journaltitle = {Physical Review E},
  shortjournal = {Phys. Rev. E},
  volume = {72},
  pages = {056218},
  issn = {1539-3755, 1550-2376},
  doi = {10.1103/PhysRevE.72.056218},
  url = {https://link.aps.org/doi/10.1103/PhysRevE.72.056218},
  urldate = {2020-12-27},
  langid = {english},
  number = {5}
}

@article{yuk2011shape,
  title={Shape memory alloy-based small crawling robots inspired by C. elegans},
  author={Yuk, Hyunwoo and Kim, Daeyeon and Lee, Honggu and Jo, Sungho and Shin, Jennifer H},
  journal={Bioinspiration \& biomimetics},
  volume={6},
  number={4},
  pages={046002},
  year={2011},
  publisher={IOP Publishing}
}

@article{zhen2015c,
  title={C. elegans locomotion: small circuits, complex functions},
  author={Zhen, Mei and Samuel, Aravinthan DT},
  journal={Current opinion in neurobiology},
  volume={33},
  pages={117--126},
  year={2015},
  publisher={Elsevier}
}
\end{document}